# Acoustic Higher-Order Topological Insulators Protected by Multipole Chiral Numbers

Yuzeng Li, Qicheng Zhang, and Chunyin Qiu[*]

* To whom correspondence should be addressed: cyqiu@whu.edu.cn

Key Laboratory of Artificial Micro- and Nano-Structures of Ministry of Education and School of Physics and Technology, Wuhan University, Wuhan 430072, China

*Abstract.* Recently, the higher-order topological phases from the chiral AIII symmetry classes are characterized by a $\mathbb{Z}$ topological invariant known as the multipole chiral numbers, which indicate the number of degenerate zero-energy corner states at each corner. Here, we report the first experimental realization of higher-order topological insulators protected by multipole chiral numbers with using acoustic crystals. Our acoustic measurements demonstrate unambiguously the emergence of multiple corner states in the middle of the gap, as predicted by the quantized multipole chiral numbers. Our study may provoke new possibilities for controlling sound, such as acoustic sensing and energy trapping.

*Introduction.*—Since the discovery of topological insulators (TIs) [1,2], tremendous effort has been devoted into the search for exotic topological phases of matter. One fundamental feature of such fascinating phases is the bulk implication of protected boundary modes, dubbed bulk-boundary correspondence. Recently, higher-order topological insulators (HOTIs) have attracted considerable interest owing to the emergence of unconventional bulk-boundary correspondence [3-12]. In comparison to conventional "first-order" topological insulators that endows *d*-dimensional topological materials with gapless states at (*d*-1) dimensions, the higher-order band topology of *d*-dimensional *n*th-order TI is characterized by gapped states at the (*d*-*n*)-dimensional boundary. Currently, there are two complementary physical mechanisms that have been proposed to realize the HOTIs: the fractional quantization of corner charge due to certain Wannier center configurations [5,7,9,10,12], and the existence of single in-gap states at corners characterized by boundary-localized mass domains [5-8,11]. Rapidly, these concepts have further been extended to semimetals systems [13-19], Floquet systems [20-23], as well as non-Hermitian systems [24-27].

Very recently, Benalcazar *et al.* demonstrate the existence of a $\mathbb{Z}$ classification for HOTIs in class AIII, and identify that *N* zero-energy states at each corner are protected by multipole chiral numbers (MCNs) [28]. The MCNs generalize the real-space representation of 1D winding number [29] to higher-dimensional systems. Comparing with the 1D winding number, the MCNs are built from sublattice multipole moment operations and apply to higher-order, boundary-obstructed topology phases. The existence of phases with MCNs reveals a richer classification of HOTIs, provides a broader understanding of boundary-obstructed topological phases beyond the Wannier center [5,7,9,10,12] and mass domain [5-8,11] perspectives, and has implications for the further classification of HOTIs in interacting systems. Even in the chiral-symmetry-preserving disorder that break translation symmetry and crystalline symmetry, the MCNs produce a quantized value. However, the HOTIs with large MCNs require increasingly stronger longer-range hoppings, these phases may be hard to attain in solid-state systems, where the electron's hoppings attenuate with separation. Due to the exceptional macroscopic controllability and tunability, the classical artificial crystals have been considered as promising platforms for studying various HOTIs. So far, there have been extensive studies of various HOTIs in classical systems [30-48], such as mechanical [30,37], photonic [32-36], electric circuits [38-39], and acoustic [40-48] systems.

Here, we first construct the HOTIs protected by MCNs and report on the observation of a sample with $N_{xy} = 4$ (MCNs in 2D) in an acoustic metamaterial. The sample is the acoustic quadrupole topological insulator (QTI) with third-range hoppings terms in the horizontal and vertical direction, and the third-range hoppings terms are emulated by long tubes between two acoustic cavities. From the measured bulk and boundary responses to local acoustic excitations, we observe four degenerate corner states at each corner, in contrast to the acoustic QTI that exhibits only one corner state. All experimental results reproduce well our full-wave simulations performed with COMSOL Multiphysics (an established commercial soft solver based on the finite element method).

*Tight-binding model.*—Benalcazar et al. [28] considered the QTI with long-range hopping terms and found that the multiple corner states are protected by MCNs. Note that the QTI with long-range hoppings holds chiral symmetry. This caused the Bloch Hamiltonian of the system in momentum space is described as follows:

$$H(\mathbf{k}) = \begin{pmatrix} 0 & Q \\ Q^\dagger & 0 \end{pmatrix} \quad (1)$$

where $Q_{11} = -Q_{22}^* = -t_0 - t_1 e^{-ik_x} - t_m e^{-imk_x}$, $Q_{12} = Q_{21}^* = t_0 + t_1 e^{ik_y} + t_m e^{imk_y}$, $t_0$ and $t_1$ are hopping amplitudes within and between unit cells, $t_m$ is long-range hopping amplitude among the *m*th nearest-neighbor unit cells in the horizontal and vertical direction. Figure 1(a) shows a schematic illustration of the model for $m = 2$. According to the real-space formula for calculating the MCNs of 2D chiral-symmetric systems:

$$N_{xy} = \frac{1}{2\pi i} Trlog\left(\bar{Q}_{xy}^A \bar{Q}_{xy}^{B\dagger}\right) \in \mathbb{Z} \quad (2)$$

here $\bar{Q}_{xy}^S = U_S^\dagger Q_{xy}^S U_S$ is the sublattice multipole moment operators projected into the spaces $U_S$, sublattice $S = A, B$. Consider a lattice in 2D with $L_j$ unit cells along the $j$ direction ($j = x, y$). The sublattice multipole moment operators $Q_{xy}^S = \sum_{\mathbf{R}, \alpha \in S} |\mathbf{R}, \alpha\rangle Exp\left(-i\frac{2\pi xy}{L_x L_y}\right) \langle \mathbf{R}, \alpha|$, each unit cell is labeled by $\mathbf{R} = (x, y)$. Thus, $N_{xy}$ for the systems with different parameters is shown in Fig. 1(b). Figs. 1(c)-(f) show the correspondence between $N_{xy}$ and corner states for the model with fixed $t_1/t_0 = 2.5$. For $N_{xy} = 1$, the system possesses a bulk band gap around zero energy and both the quadrupole moment [3,4,28]. Starting from this phase and increasing $t_2/t_0$, a bulk band-gap-closing phase transition occurs at $t_2/t_0 = 1.5$, after which topological invariant now shows that this system is in a nontrivial phase with $N_{xy} = 4$. We can clearly see the energy spectrum of the open boundary condition in the case of two-phase transition ($N_{xy} = 1 \rightarrow 4$), shown in Fig. 1(d), which leads to changes in the number of corner-localized states. The model possesses four degenerate corner states but a single zero energy state localized to each of its corners in the case of $N_{xy} = 1$ [Fig. 1(e)]. When $N_{xy} = 4$, there are four degenerate corner states localized at each corner, shown in Fig. 1(f). Therefore, the number of topological corner states at zero energy is then given by $n = 4N_{xy}$ according to the bulk-boundary correspondence. To avoid the finite size effect, the size is taken as 50×50 cells in the process of calculating $N_{xy}$.

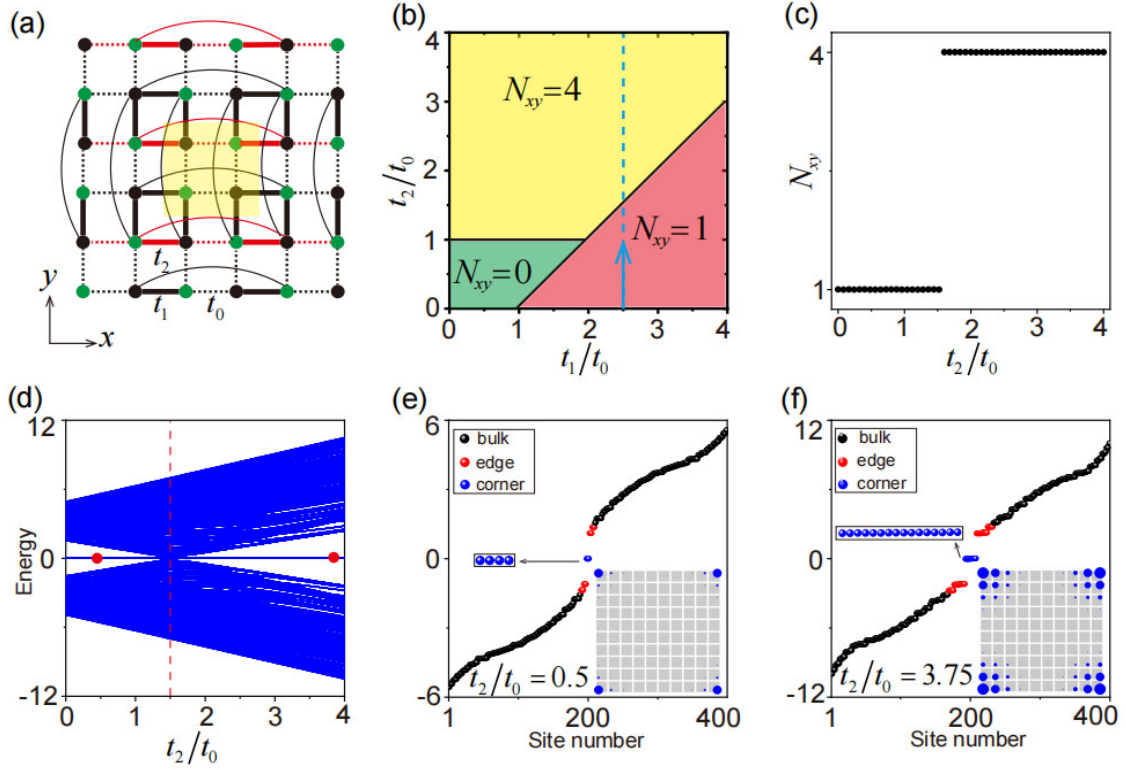

FIG. 1. Tight-binding model. (a) Lattice model for a QTI with third-neighbor hoppings. (b) Associated phase diagrams. (c) MCNs calculated as a function of $t_2/t_0$, indicated as the bule line in (b). (d) The energy spectrum for this system for fixed $t_1/t_0 = 2.5$. (e) and (f) are the energy levels marked red in (d). And the probability density of the corresponding zero energy corner states in the inset. (d)-(f) The system size was 10 × 10 cells.

The calculations so far show that the $N_{xy}$ can take a largest value equal to 4 in the presence of third-neighbor hoppings. By considering a fifth-neighbor hopping term ($m = 3$), we were able to generate topological phases with $N_{xy} = 9$ (See *Supplemental Material*), and this leaves little doubt that $N_{xy}$ can take larger integer value when $m$ is a larger value. In 3D, chiral-symmetric third-order phases are characterized by distinct integer values of $N_{xyz}$ (MCNs in 3D), which indicate the number of degenerate states localized at each corner in the 3D structure (See *Supplemental Material*). In addition, the phases with nonzero MCNs are known as boundary-obstructed topological phases, which supports that phase transitions between phases with different MCNs need not close the bulk band gap but, at a minimum, must close some lower-dimensional edge or surface band gap. Even hinge-obstructed topological phases may appear in 3D structures (See *Supplemental Material*).

*Acoustic realizations.*—This model can be implemented with cavity-tube structures in acoustic systems. The cavity resonators emulate atomic orbitals, the narrow tubes introduce hoppings between them. Therefore, we can directly introduce a long tube between the two cavity resonators to emulate the long-range hopping. Thus, we found that the sign of hopping is tunable via adjusting the length of the narrow tubes (See *Supplemental Material*). As we all known, the positions of the narrow tube can control the hopping signs [46,48] to achieve acoustic QTI [Fig. 2(d)]. As a result, we elaborately design long tube on acoustic QTI to realize this phase with $N_{xy} = 4$, shown in Fig. 2(a). The structures provide effectively the onsite energy $\sim 5.7$ kHz, $t_0 = t_1 \approx 15$ Hz, and $t_2 \approx 90.3$ Hz for Fig. 2(a). In Fig. 2(d), $t_0 \approx 20.3$ Hz and $t_1 \approx 110.5$ Hz (details in *Supplementary Material*).

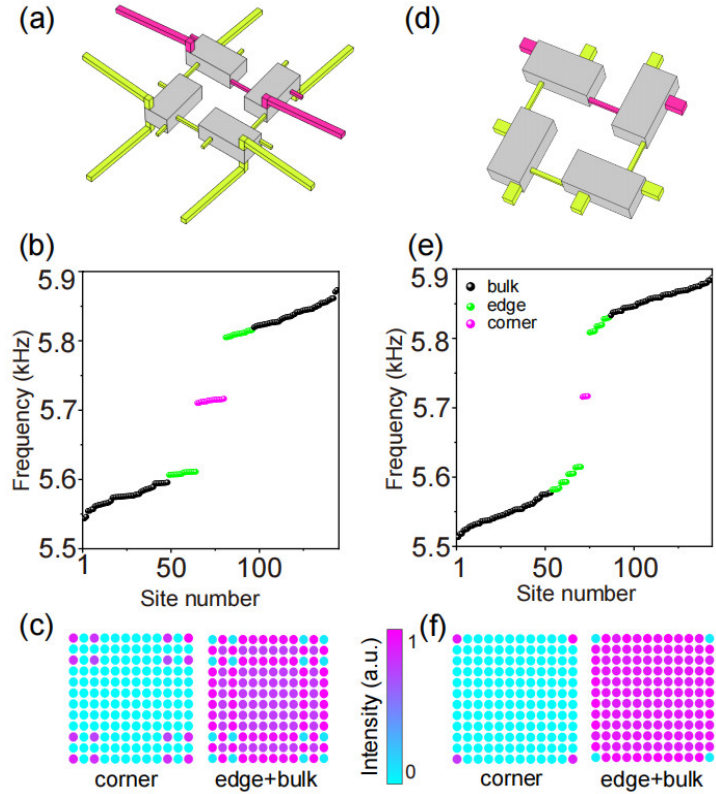

FIG. 2. Acoustic realization of the MCNs protected HOTI. (a)-(c) Unit-cell geometry of the 2D acoustic model with $N_{xy} = 4$, eigenvalue spectrum for a finite-sized sample, and averaged intensity distributions of the bulk and corner states extracted for sample. (d)-(f) Similar to (a)-(c), but for the 2D acoustic QTI with $N_{xy} = 1$.

Figures. 2(b) and (e) provide the eigenfrequency spectra simulated for the above samples of finite lattices (6×6 cells). It is observed that corner states appear in the middle of the gap. And the number of corner states in Fig. 2(b) is four times that in Fig. 2(e), which is consistent with the results of the tight-binding predictions. As can be seen in Fig. 2(c), the corner states with phase of $N_{xy} = 4$ appear in the four acoustic cavities (same sublattice) at each corner in the case of $t_2 \gg t_0$ and $t_2 \gg t_1$. This is because the tight-binding model holds separability [28,49-51], which implies that the corner states given by $\phi_x \otimes \phi_y$. $\phi_{x(y)}$ are topological boundary states of separated 1D model (see *Supplementary Material*). As a contrast, the corner states of the acoustic QTI naturally only appear in the acoustic cavities at the corners [Fig. 2(f)]. These facts can help us distinguish well that the multiple corner states with $N_{xy} = 4$ and the acoustic QTI with $N_{xy} = 1$ in the following experiments. It is noteworthy that only the corner states are topologically protected, whereas the edge states are not.

*Experimental results.*—Figure 3(a) shows our experimental sample for the 2D acoustic model with $N_{xy} = 4$. It consists of 12×12 acoustic cavities in the *x* and *y* directions. The experimental samples are fabricated by 3D printing with a photosensitive resin material. Both the input and output signals were recorded and frequency-resolved with a multi-analyzer system (B&K Type 4182). On each cavity, two small holes were perforated for inserting sound source or probe, and they were sealed when not in use. Figure. 3(b) presents the Fourier spectrum (color scale) performed for the experimental sound signals in time-space domain. It shows a good agreement with predicted bulk band structure (red line). We have calculated the spatial distributions of the corner states in Fig. 2(c), so we focus on the corresponding four cavities at each corner (1-4 marked in Fig. 3(a)). We then plot the intensity spectra for the four cavities, as shown in magenta curve in Fig. 3(c). One can see four intensity spectra of the corner states and average intensity spectra for bulk states, which is consistent with the simulation results [Fig. 3(d)]. To further characterize the multiple corner states, we plot the spatial distributions of the acoustic response integrated over frequency ranges corresponding to each type of spectral peak (the three different integration regions are indicated in Fig. 3(c)). The corner states are clearly observable, but the bulk states cannot be recognized because the intensity is roughly the same as that of the corner states. The experimental field distributions are highly consistent with the simulated results presented in Fig. 3(f).

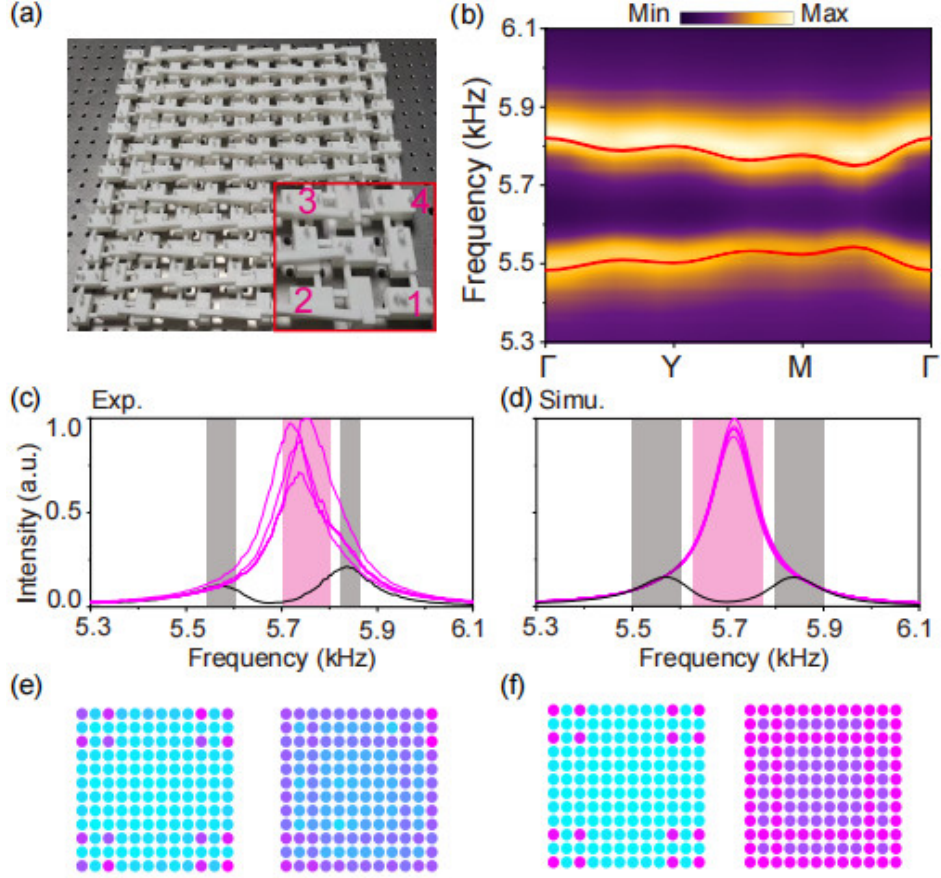

FIG. 3. Experimental observation of the phase $N_{xy} = 4$. (a) Photograph exemplified for the sample. Inset: a local view of a corner of sample. (b) Experimentally measured (color scale) and theoretically predicted (red line) bulk spectra. (c)-(d) Measured and simulation results of intensity spectra for four acoustic cavities (magenta line) and average intensity spectra of bulk states (black line). (e)-(f) The experiment (e) and simulation (f) of spatial distributions of the acoustic response intensity, integrated over different sets of frequencies corresponding to the corner and bulk spectra.

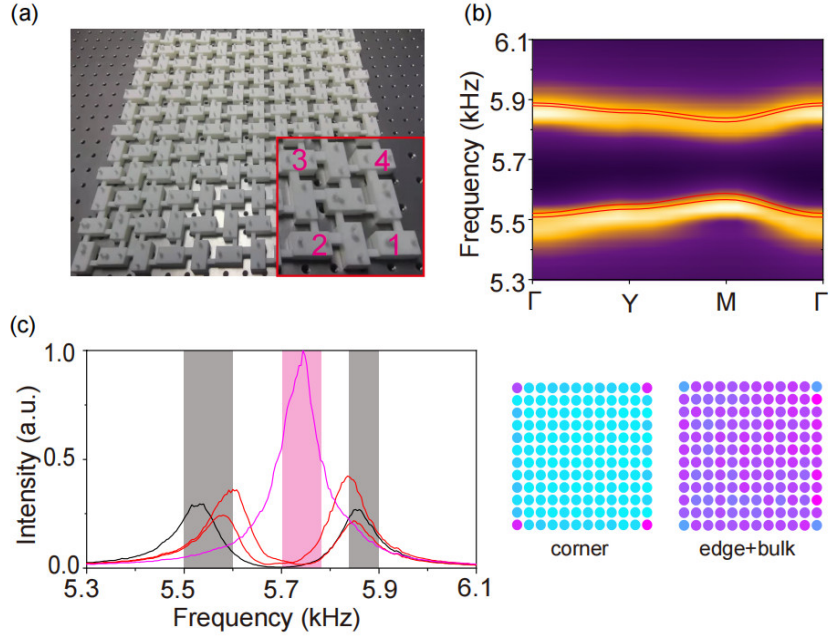

FIG. 4. Experimental measurements of the phase $N_{xy} = 1$. (a) A photo of acoustic metamaterial. Inset: an enlarged view of a corner. (b) Measured bulk spectra (color scale) excellently match the simulation results (red line). (c) Left: Intensity spectra measured for the corner states marked 1, edge states marked 2 and 4, and bulk states marked 3. Right: Spatial distributions of the acoustic response intensity, integrated over the spectral peaks in left.

Similar measurements were conducted on the samples with $N_{xy} = 1$. Naturally, there is only one intensity spectra of corner states (cavity marked 1 in Fig. 4(a)) for the model, and the spatial distributions of the acoustic response intensity also implies that there is a single zero-energy state at each corner [Fig. 4(c)]. The edge and bulk states are closer in frequency, leading to overlaps in the integrated intensity spectra. These results match the predictions in Fig. 2 and show that the MCNs really predict the number of degenerate zero-energy states localize at each corner of a system.

*Conclusions.*—Following the theory of MCNs that describes the number of degenerate states localized at each corner, we designed and fabricated acoustic samples with different MCNs, and experimentally validated the numbers of mid-gap states and measured intensity spectra for corner states are consistent with the theoretical analysis. Note that the HOTIs with large MCNs require increasingly stronger longer-range hoppings, which are emulated by long tubes between two acoustic cavities. Therefore, acoustic samples with larger MCNs can be ideally designed in 2D and 3D (see *Supplementary Material*). In addition, comparing to single corner state localized at each corner, the multiple corner states have stronger control ability of sound energy concentration. And our results can be extended to topological semimetals with multi hinge states through 3D stacking [13,16-17]. Potential applications can be anticipated for such unique topological corner states, e.g., acoustic sensing and energy trapping.

## Acknowledgements

This work is supported by the National Natural Science Foundation of China (Grant No. 11890701, 12104346, 11674250), the Young Top-Notch Talent for Ten Thousand Talent Program (2019-2022).

*Supplementary Information for*

# Acoustic Higher-Order Topological Insulators Protected by Multipole Chiral Numbers

[1]Key Laboratory of Artificial Micro- and Nano-Structures of Ministry of Education and School of Physics and Technology, Wuhan University, Wuhan 430072, China

## Supplementary Note 1.
## Realization Of Long-Range Hoppings

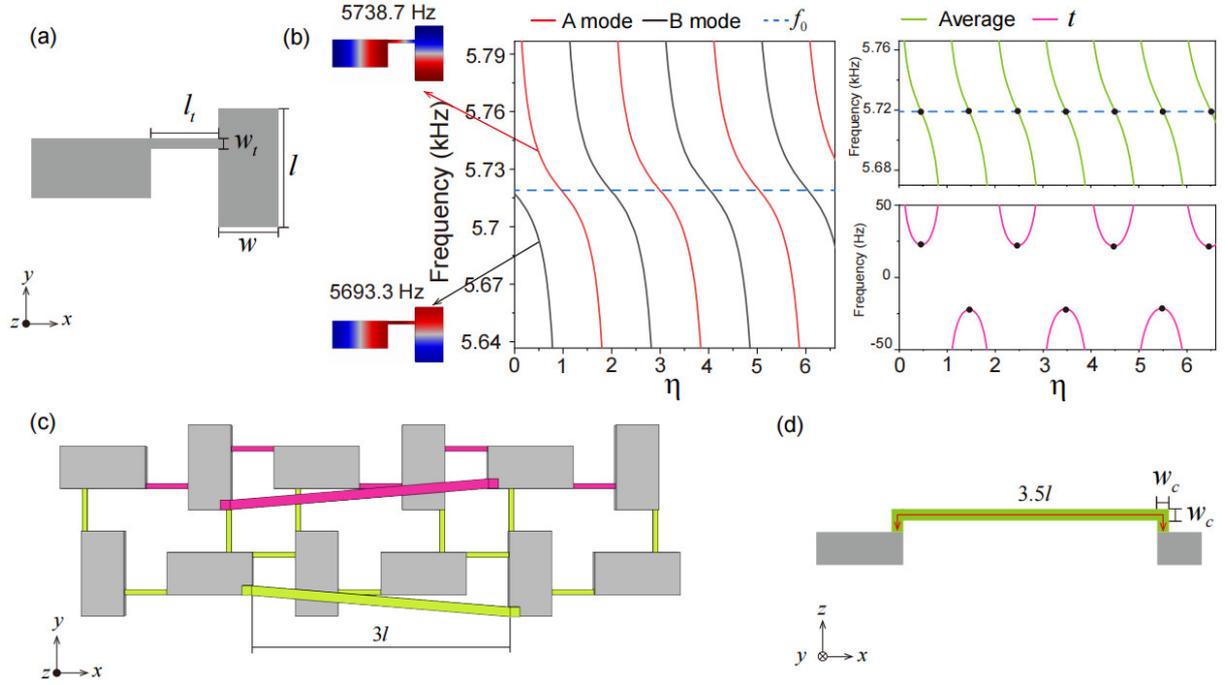

FIG. S1. Acoustic realization of long-range hoppings. (a) Two coupled-resonator systems. (b) Left: Frequencies of the coupled modes A and B as function of the parameter $\eta = l_t/l$. Right: Effective hopping $t$ and average of the two split eigenfrequencies plotted as a function of $\eta$. (c) An $x$-chain consisting of three acoustic quadrupole cells, where the connecting tubes tied to positive and negative couplings are marked in green and magenta, respectively. (d) Projection view of two cavities with long-range hopping in $xz$ coordinate system.

Now we consider the influence of length $l_t$ of tube in the cavity-tube structures, the two-cavity system is shown in Fig. S1(a). Each cavity has a size of 30mm×15mm×11mm, which supports a fundamental dipole resonance mode at $f_0 = 5716.7$ Hz. The square tubes of widths $w_t = 1.8$ mm. Throughout this work, air density $\rho = 1.29\ kg/m^3$ and sound speed $c = 343\ m/s$ are used for full-wave simulations, performed with the commercial software COMSOL Multiphysics (Pressure Acoustics Module). The system can be modeled by a two-level Hamiltonian, $H = \begin{pmatrix} f_0 & t \\ t & f_0 \end{pmatrix}$, where the magnitude of hopping $t$ is tunable by the width of narrow tube [1]. The composed system exhibits split eigenfrequencies $f_\pm = f_0 \pm t$ associated with eigenvectors $(1/\sqrt{2})(1, \pm1)^T$: the signs $+$ and $-$ characterize the modes formed by in-phase and out-of-phase couplings, respectively. The in-phase and out-of-phase couplings are represented by A and B mode, respectively. Thus, the A mode is at a higher (lower) frequency, indicating that $t$ is positive (negative). The eigenfrequencies of the two-cavity system are functions of $\eta = l_t/l$ is shown in Fig. S1(b). As $\eta$ approaches integers, sign of the hopping $t$ is reversed. Meanwhile, the green line and $f_0$ intersect near where $\eta$ equals a half-integer. That implies the model can be regarded as an ideal tight-binding model in this case.

Therefore, we use long tube to achieve long-range hoppings in the acoustic QTI. Firstly, to accurately realize the tight-binding model of quadrupole in acoustics, $w$ and $l_t$ are equal to $l/2$. Then, in Fig. S1(c), the shortest distance between two cavities with third-neighbor hopping is $3l$. We must choose the length of the long tube is $3.5l$ in Fig. S1(d), and $w_c = 3.4$ mm. However, when $\eta = 3.5$, the sign of hopping is opposite to that when $\eta = 0.5$. We know, relocating the connecting narrow tube to the other side of the dipole mode node line can also result in the inversion of the hopping's sign [1,2]. Thus, we change the position of the long tube to keep consistent with the sign of the nearest-neighbor hoppings in Fig. S1(c). We can design the long-range hopping in the $y$ direction through the same way. The long tubes in the $y$ direction can be placed on the back of the acoustic cavity to avoid crossing with the long tubes in the $x$ direction.

## Supplementary Note 2.
### Acoustic Designs Of The Model

Figure. S2(a) shows unit-cell geometries of acoustic QTI. The lattice constants are $a = 75$ mm, and the square tubes of widths $w_0 = 1.8$ mm and $w_1 = 4.6$ mm correspond to the couplings $t_0$ and $t_1$ in the tight-binding model. The unit-cell of acoustic QTI with third-neighbor hoppings is shown in Fig. S2(c). The lattice constant does not change, and half of the long tube was taken to calculate the bulk band structure. For simplicity, the $w_0 = w_1 = 1.8$ mm, which implies $t_0 = t_1$ in the tight-binding model. And $w_2 = 3.4$ mm is $t_2$ in the tight-binding model.

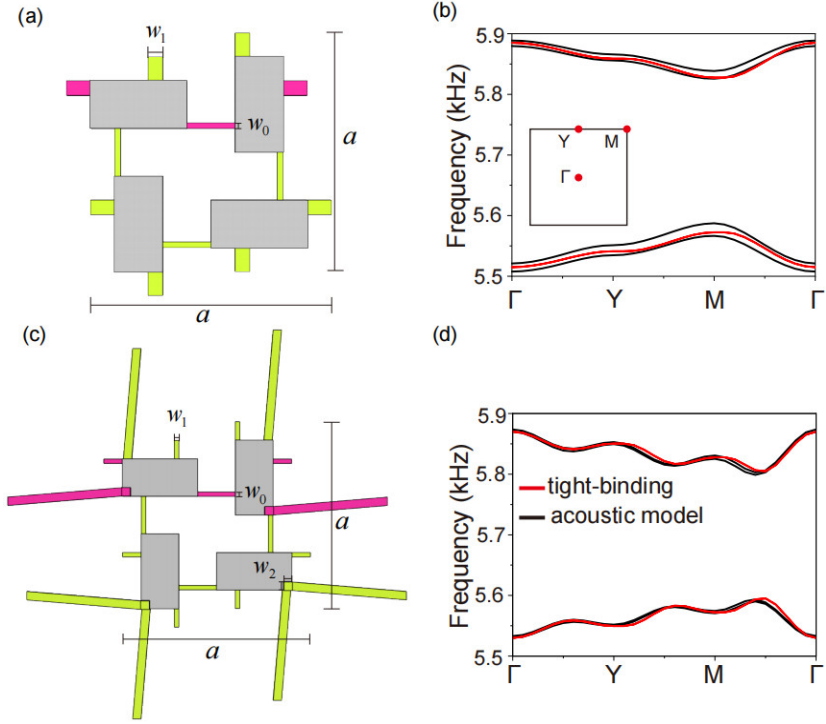

FIG. S2. Acoustic constructions of the model. (a), (c) Unit-cell geometry of the acoustic QTI and the model with third-neighbor hoppings. (b), (d) Bulk band structure of the corresponding acoustic model (black line) and tight binding model (red line).

We fit the effective parameters (onsite energy and hoppings) based on the bulk dispersion simulated with COMSOL Multiphysics, in which high-symmetry points are particularly taken into accounted. Figures. S2(b) and (d) show the bulk band structures of the acoustic QTI and the model with third-neighbor hoppings, respectively. Notice that the numerical bulk bands are not strictly twofold degenerate as predicted by theory, due to the presence of unavoidable long-range and inter-mode couplings that weakly break the symmetries in the real structures. These unwanted effects can be reduced through an optimization process [2]. To fit the bulk dispersion, we consider the average frequency of the split bands. We fit the onsite energy $f_0 = \frac{1}{3}(f_\Gamma + f_Y + f_M)$, where $f_\Gamma$, $f_Y$ and $f_M$ are average frequencies of the conduction and valence bands at the Γ, Y and M points, respectively. For acoustic QTI, the hoppings based on the relations $2\sqrt{2}(t_0 + t_1) = \Delta\Gamma$ and $2\sqrt{2}(t_1 - t_0) = \Delta M$. And the fitting process leads to $f_0 \approx 5.7013$ kHz, $t_0 \approx 20.3$ Hz and $t_2 \approx 110.5$ Hz for Fig. S2(b). For the model in Fig. S2(c), the hoppings based on the relations $2\sqrt{2}(t_0 + t_2) = \Delta\Gamma$ and $2\sqrt{2}(2t_0^2 + t_2^2 + 2t_0 t_2) = \Delta Y$, and leads to $f_0 \approx 5.7011$ kHz, $t_0 \approx 15$ Hz and $t_2 \approx 90.3$ Hz. The ΔΓ, ΔY and ΔM are respectively the band gaps at the Γ, Y and M points. The bulk band structure (red line) in Fig. S2(d) is highly consistent with the results of COMSOL Multiphysics, which confirms the effectiveness of the fitting parameters.

## Supplementary Note 3.
### Zero-energy corner states of Model

The separated 1D model is the Su-Schrieffer-Heeger (SSH) with the third-neighbor hoppings. The Hamiltonian of the finite system with $N$ cells is given by

$$H = H_A + H_B$$

$$H_A = t_0 \sum_{n=1}^{N} |A,n\rangle\langle B,n| + t_1 \sum_{n=2}^{N} |A,n\rangle\langle B,n-1| + t_2 \sum_{n=3}^{N} |A,n\rangle\langle B,n-2|$$

$$H_B = t_0 \sum_{n=1}^{N} |B,n\rangle\langle A,n| + t_1 \sum_{n=1}^{N-1} |B,n\rangle\langle A,n+1| + t_2 \sum_{n=1}^{N-2} |B,n\rangle\langle A,n+2|$$

Note that the Hamiltonian have the chiral symmetry. Because of this property, the Hamiltonian can decouple into two interpenetrating sublattices, $A$ and $B$ coupled to each other but not within themselves. And the zero modes can be decomposed into left- and right-handed chirality states. In the thermodynamical limit $N \to \infty$, the zero energy boundary states can also be calculated exactly in the absence of translational invariance. The zero-energy boundary states should satisfy the eigenvalue equation $H|\phi\rangle = 0$, where $|\phi\rangle = a_n|A,n\rangle + b_n|B,n\rangle$ is the eigenstate of the zero energy. This gives us $2N$ equations for the amplitudes $a_n$ and $b_n$, which read

$$t_0 a_n + t_1 a_{n+1} + t_2 a_{n+2} = 0$$
$$t_0 b_n + t_1 b_{n-1} + t_2 b_{n-2} = 0$$

The boundary equations are $t_0 a_{N-1} + t_1 a_N = 0, t_0 a_N = 0$ and $t_0 b_2 + t_1 b_1 = 0, t_0 b_1 = 0$. Since the Hamiltonian have the inversion symmetry, which mean that the amplitude $b_n$ can be obtained accordingly by labeling the indexes of the unit cells from the opposite direction. For the zero energy boundary states of the left 'odd' sublattice, we consider semi-infinite geometry, and the transfer matrices can be expressed as

$$\begin{pmatrix} a_{n+2} \\ a_{n+1} \end{pmatrix} = \begin{pmatrix} -\frac{t_1}{t_2} & -\frac{t_0}{t_2} \\ 1 & 0 \end{pmatrix} \begin{pmatrix} a_{n+1} \\ a_n \end{pmatrix}$$

Defining this equation as $\boldsymbol{A}_{n+1} = S\boldsymbol{A}_n$. We can diagonalize the transfer matrix $D = U^{-1}SU$ with

$$D = \begin{pmatrix} \lambda_- & 0 \\ 0 & \lambda_+ \end{pmatrix}, U = \begin{pmatrix} \lambda_- & \lambda_+ \\ 1 & 1 \end{pmatrix}, U^{-1} = \frac{1}{\lambda_- - \lambda_+}\begin{pmatrix} 1 & -\lambda_+ \\ -1 & \lambda_- \end{pmatrix}$$

and $\lambda_\pm = \left(-t_1 \pm \sqrt{t_1^2 - 4t_0 t_2}\right)/2t_2$. These candidates zero energy boundary modes must be normalizable which requires $|\lambda_\pm| < 1$. Then $-1/\ln|\lambda_\pm|$ gives the localization length of the edge mode. For the case $t_1, t_2 > 0$ and $t_0 = 1$. When $t_2 > t_1 - 1$ and $t_2 > 1$, one can see that $|\lambda_\pm| < 1$. This means that in these regions that are indicated by winding number $\nu = 2$ the system has two independent zero mode solutions. By solving the recurrence formula $S^n = UD^n U^{-1}$, we get

$$a_n = \frac{a_1(-\lambda_+ \lambda_-^n + \lambda_- \lambda_+^n) + a_2(\lambda_-^n - \lambda_+^n)}{\lambda_- - \lambda_+}$$

There are two independent zero mode solutions in case of $t_2 \to \infty$. And, the two zero energy boundary states are mainly localized at the left edges $A_1$ and $A_2$. The two edge states can have the same localization length since $\sqrt{t_1^2 - 4t_0 t_2}$ can be purely imaginary. Naturally, the corner states $\phi_x \otimes \phi_y$ within a corner mainly occupied the four sites (same sublattice) in the case of $t_2 \gg t_0$ and $t_2 \gg t_1$.

*Supplementary Note 4.*

*Acoustic HOTIs with Larger MCNs in 2D Model*

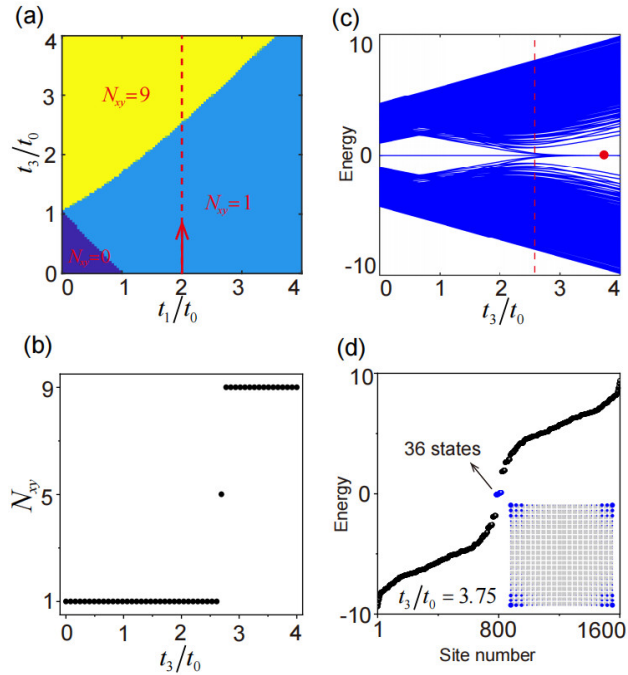

FIG. S3. Tight-binding model with larger MCNs in 2D. (a) Phase diagram versus the coupling ratios $t_1/t_0$ and $t_3/t_0$. (b)-(c) Eigenspectrum and $N_{xy}$ vs $t_3/t_0$ calculated for a finite system with 60×60 cells, indicated as the red line. (d) are the energy levels marked red in (b). The probability density of the corresponding 36 states in the inset.

Here we present some numerical details for HOTIs protected by larger $N_{xy}$ not discussed in detail in our main text. By considering fifth-neighbor hopping term ($t_3$), the model could support topological phases with $N_{xy} = 9$, and this leaves little doubt that $N_{xy}$ can take larger integer value when $m$ is larger. There exist topological phases characterized by $N_{xy} = 0,1,9$ in this model when $m = 3$. Thus, the $N_{xy}$ for systems with different parameters is shown in Fig. S3(a). As $t_3/t_0$ is increased from 0 to 4, this system undergoes two separate phase transitions by fixing $t_1/t_0 = 2$. The phase transitions in Fig. S3(b), $N_{xy} = 1 \to 9$, result in changes in the number of corner-localized states is shown in Fig. S3(d).

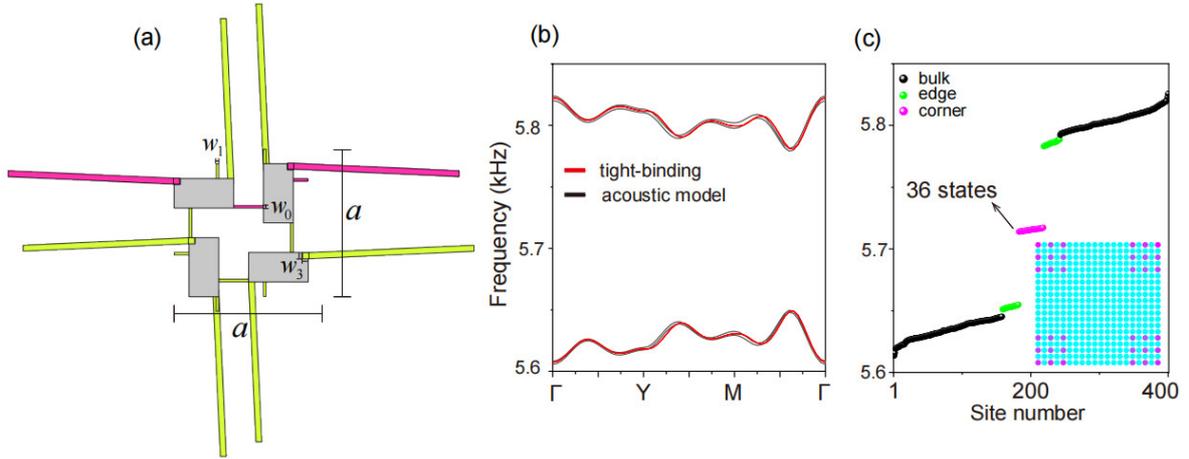

FIG. S4. Acoustic realization of the phase $N_{xy} = 9$. (a) Unit-cell geometry of the acoustic model. (b) Bulk band structure of the corresponding acoustic model (black line) and tight binding model (red line). (c) Simulated eignfrequencies of the system with 10×10 cells. Inset shows the intensity distributions of the corner states.

We use coupled acoustic cavities to implement this phase $N_{xy} = 9$ based on the tight-binding model. The lattice constants are $a = 75$ mm, $w_0 = w_1 = 1.4$ mm and $w_3 = 3.1$ mm, which implies that $f_0 \approx 5.715$ kHz, $t_0 = t_1 \approx 8.2$ Hz and $t_3 \approx 59.7$ Hz through the method of fitting parameters in *Supplementary Note 2*. The bulk band structure (red line) in Fig. S4(b) is highly consistent with the results of COMSOL Multiphysics, which confirms the effectiveness of the fitting parameters. As we can imagine, the shortest distance between two cavities with fifth-neighbor hopping is $5.5l$. We must choose the length of the long tube is $6.5l$ in Fig. S4(a). Therefore, when $\eta = 6.5$, the sign of hopping is equal to that when $\eta = 0.5$ in Fig. S1(b). The long tube located at same sides of the cavity's nodal line with the short tube, shown in Fig. S4(a). Figure S4(c) provide the eigenfrequency spectra simulated for the samples of finite lattices. It is observed that 36 corner states emerge inside the mid-gap and the field distributions of the corner states. These facts provide compelling evidence for the theory of MCNs and the effectiveness of constructing long-range hoppings in acoustic cavities.

## Supplementary Note 5.
### Acoustic HOTIs with MCNs in 3D Model

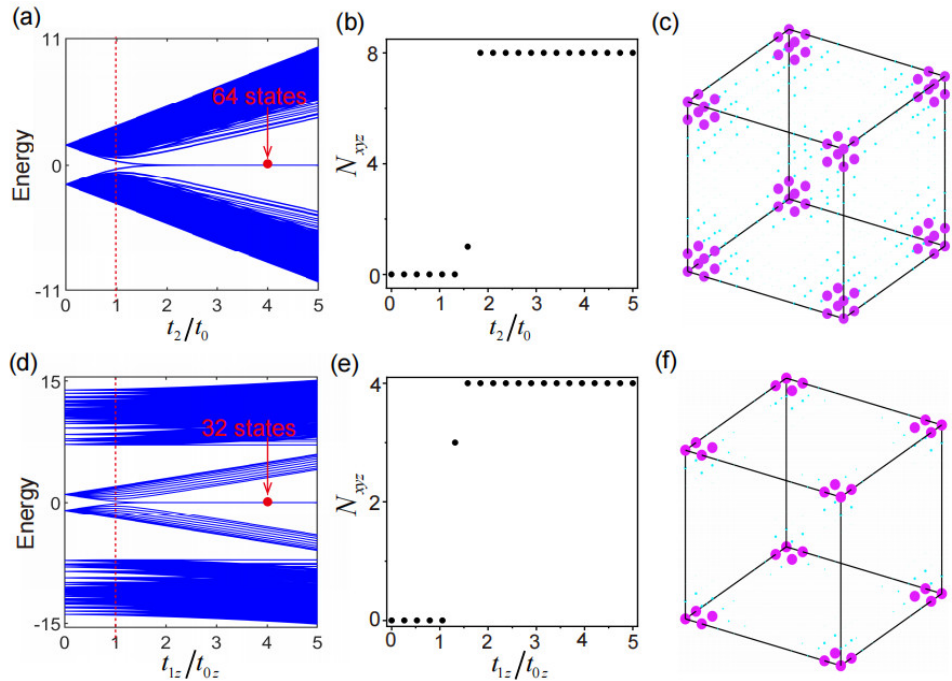

FIG. S5. Correspondence between the appearance of zero energy states, change in the topological invariant $N_{xyz}$, and the probability density of the corner states. (a)-(b) Number of states relative to their energy and $N_{xyz}$ as a function of $t_2/t_0$ calculated using open boundary conditions. (c) The probability density of corner states marked by red in (a). (d)-(f) Similar to (a)-(c), but for the phase $N_{xyz} = 4$, $t_{0,x/y}/t_{0z} = t_{1,x/y}/t_{0z} = 1$ and $t_{2,x/y}/t_{0z} = 8$. In all simulations of both systems, the system size was 15×15×15 cells.

In 3D, chiral-symmetric third-order phases are characterized by distinct integer values of $N_{xyz}$, which indicate the number of degenerate states localized at each corner in the 3D structure. $N_{xyz}$ is integer topological invariant of MCNs in 3D. The $N_{xyz}$ is given by [3]

$$N_{xyz} = \frac{1}{2\pi i} Tr\log\left(\bar{O}^A_{xyz} \bar{O}^{B\dagger}_{xyz}\right) \in \mathbb{Z}$$

here $\bar{O}^S_{xyz} = U_S^\dagger O^S_{xyz} U_S$, for $S = A, B$, is the sublattice multipole moment operators projected into the spaces $U_S$, Consider a lattice in 3D with $L_j$ unit cells along the $j$ direction ($j = x, y, z$). The sublattice multipole moment operators $O^S_{xyz} = \sum_{R,\alpha \in S} |R, \alpha\rangle Exp\left(-i\frac{2\pi xyz}{L_x L_y L_z}\right)\langle R, \alpha|$, each unit cell is labeled by $R = (x, y, z)$. Thus, the octupole insulator is a simple model with $N_{xyz} = 1$ in 3D [2]. Here, we consider the third-neighbor hopping terms $t_2$ in the octupole insulator, $t_0$ and $t_1$ are hopping amplitudes within and between unit cells of octupole insulator. Figures. S5(a) and (b) present numerical simulations of energy spectrum and $N_{xyz}$ with $t_1/t_0 = 0$. As can be seen, the closing of the bulk band gap near $t_2/t_0 = 1$ coincides with 32 states departing from each of the upper and lower bulk bands and becoming pinned at $\varepsilon = 0$. Note the bulk band gap is not strictly closed at $t_2/t_0 = 1$ due to the finite size effect. These 64 states in total are those corner-localized states shown in Fig. S5(c), which are predicted to exist as $N_{xyz} = 8$ for $t_2/t_0 > 1$.

Here, we only consider the third-neighbor hopping terms $t_{2,x/y}$ in the $x$ and $y$ direction, and $t_{2,x/y} \gg t_{0,x/y}$, $t_{2,x/y} \gg t_{1,x/y}$. The $t_{0,x/y/z}$ and $t_{1,x/y/z}$ the nearest-neighbor hoppings within a unit cell and between adjacent unit cells, respectively. Figure S5(d) presents the energy spectra as a function of the $t_{1z}/t_{0z}$. As the adiabatic growth of $t_{1z}/t_{0z}$, the z-directed hinge states close and reopen, and the xy-surfaces corner states emerge as $t_{1z}/t_{0z} > 1$. This phenomenon is a hinge-obstructed topological phase [4]. The phase transition in Fig. S5(e), $N_{xyz} = 1 \to 4$, and result in changes in the number of corner-localized states. Thus, the xy-corner states are protected by $N_{xyz} = 4$ in Fig. S5(f).

21 / 23

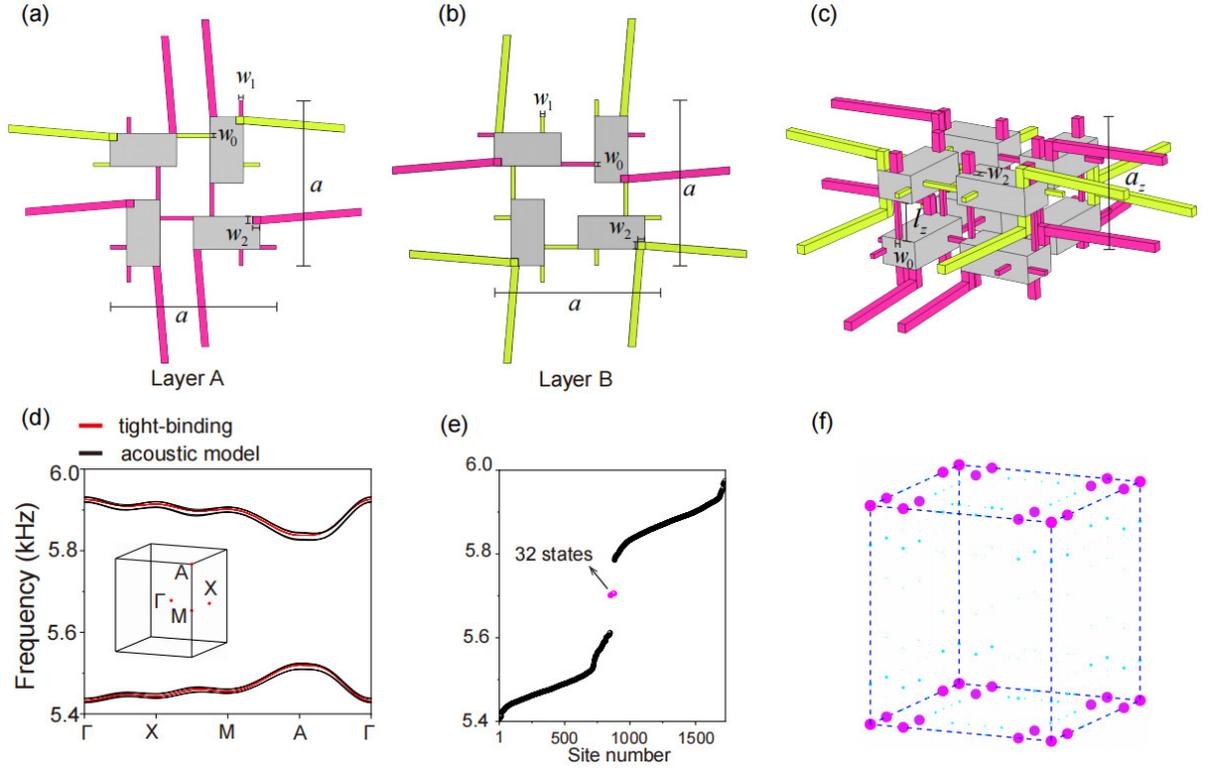

FIG. S6. Acoustic constructions of the phase $N_{xyz} = 4$. (a)-(b) In-plane geometry of the unit cell in 3D model. (c) Unit cell geometry of the layer-stacked 3D structure. (d)-(e) Bulk dispersions of the designed acoustic lattice along high symmetry lines and simulated eigenfrequencies of the system with 6×6 cells, respectively. (f) The intensity distributions of the 32 states in (e).

The tight-binding model for an octupole topological insulator is built from two quadrupole topological insulator with opposite coupling settings along the z direction. We first design the third-neighbor hoppings in two layers as shown in Figs. S6(a) and (d). These two layers are coupled along z direction with dimerized coupling strengths, resulting in the acoustic structure with $N_{xyz} = 4$. Figure S6(c) shows the unit cell geometry of acoustic model with $N_{xyz} = 4$. The lattice constants are $a = 75$ mm and $a_z = 52.2$ mm, which ensures a dipole resonance (along the length direction) of frequency ~5.6787 kHz (i.e., onsite energy) far away from the other undesired cavity modes. The length of the coupling tube in the z direction is $l_z = 15.1$ mm. The widths of the coupling tubes are fixed to be $w_0 = 1.8$ mm and $w_2 = 3.4$ mm, yielding effectively the coupling strengths $t_{0,x/y} = t_{1,x/y} \approx 17.3$ Hz, $t_{2,x/y} \approx 63.4$ Hz, $t_{0z} \approx 36.3$ Hz and $t_{1z} \approx 167.7$ Hz in the tight-binding model, as shown in Fig. S6(d). Figures S6(c)-(d) provide the eigenfrequency spectra simulated for the above three samples of finite lattices and field distributions of the xy-surfaces corner states.

These results are consistent with the theoretical prediction, and provide a practical evidence for the theory of MCNs in 3D.